\documentclass[aps,prd,amsmath,twocolumn,nofootinbib,longbibliography]{revtex4-1}

\usepackage{graphicx,amsmath,latexsym,color} 
\usepackage{subcaption}

\def\be{\begin{equation}}
\def\ee{\end{equation}}

\newcommand{\eq}[1]{Eq.~(\ref{#1})}

\newcommand{\boldtau}{\mbox{\boldmath$\tau$}}
\def\bea{\begin{eqnarray}}
\def\eea{\end{eqnarray}}\def\a{\alpha}
\def\g{\gamma}\def\t{\tau}\def\l{\lambda}\def\o{\omega}

\def\la{\langle}\def\ra{\rangle}\def\d{\delta}\def\bfr{{\bf r}}\def\k{\kappa}\def\b{\beta}
\def\o{\omega}
\def\r{\rho}

\def\D{\Delta}
\def\L{\Lambda}
\def\w{\omega}\def\O{\Omega}
\def\SRC{\rm{ SRC}}\def\bfr{{\bf r}}\def\bfs{{\bf s}}
 
 \def\l{\lambda}
\def\OC{\widehat{\cal O}_C(v)}
\begin{document}

\title{Strong Interaction  Dynamics and Fermi $\beta$ decay in the nucleon and the nucleus}
%\title{Nucleon-Nucleon Short-Ranged Correlations,   $\beta$ Decay and the Unitarity of the CKM Matrix  }

\author{Gerald A. Miller}
 
\affiliation{
University of  Washington, Seattle  \\
Seattle, WA 98195-1560}

\date{\today}

\begin{abstract}
Nuclear super-allowed $\beta$ decay has been used to obtain tight limits on the value of the CKM matrix element $V_{ud}$  that is important for unitarity  tests,
 and therefore  important for tests of the standard model.  Current requirements on precision are so intense that effects  formerly thought too small to matter are now relevant. This article is   a brief  review of personal efforts to include the effects of strong interactions on Fermi $\beta $ decay. 
First I examine the role of isospin violation in the decay of the neutron.
% that causes the value of the decay matrix element to differ from $V_{ud}$. 
The size of the necessary correction depends upon detailed strong-interaction dynamics.  The isospin violating parts of the nucleon wave function, important at the low energy of $\beta$ decay,  can be constrained by data taken at much  higher energies, via measurements, for example,  of   $ed\to e'\pi^\pm +\rm X $ reactions at Jefferson Laboratory.
 The next focus is on  the role  of nuclear short-ranged correlations, which affect the value of the  correction needed to account  for isospin violation in  extracting   the value of $V_{ud}$. The net result is that  effects previously considered as irrelevant 
have become relevant for both neutron and nuclear $\b$ decay.
\end{abstract}
 
 \maketitle

\section{Introduction}
The problem of precisely understanding $\b$ decay of nuclei  %and the neutron 
has  recently  come into prominence  because of its possible use as a probe of new physics beyond the Standard Model~\cite{Cirigliano:2019wao,Brodeur:2023eul}. Nuclear $\b$ decay transitions have been  used to extract the value of $V_{ud}$
that governs the probability that a down quark decays into an up  quark,  important for  testing the unitarity of  the CKM matrix
 \cite{Kobayashi:1973fv}. At present, the most precise determination of $V_{ud}$ is obtained from $0^+\to0^+$ (superallowed) decays of nuclei. The value  obtained from  a set of  more than 200 measurements in 21 different nuclei 
 ~\cite{Hardy:2020qwl}  is
\be V_{ud}=0.97373\pm 0.00031.\label{val}\ee
The Particle Data Group~\cite{ParticleDataGroup:2020ssz} has obtained essentially the same  central value of $V_{ud}$ but a smaller uncertainty of $\pm0.00014$. 
Extracting such values  requires that the effects of nuclear isospin breaking needs to be removed.
 This precision is remarkable, especially because the  nuclear structure theory needed to make various corrections is of very early vintage. 
 
 Concerns about the accuracy of the Towner-Hardy formalism used for the nuclear structure corrections led us to propose a different formalism \cite{Miller:2008my,Miller:2009cg} which allowed the inclusion of shell model states of higher energies.  Recent experimental work, see {\it e.g.} the review \cite{Hen:2016kwk}, showing that the effects of nucleon-nucleon short-ranged correlations can be and have been measured stimulated us 
 \cite{Condren:2022dji} to revisit the subject of nuclear superallowed $\b$ decay. We found that the effects of short-ranged correlations are relevant for superallowed $\b$ decay.
 
The subject of neutron $\b$ decay naturally arose during the study of  nuclear decays. In particular, the assumption that the quark-level $\b$ decay operator is the same as the neutron-level operator has been widely used. This assumption is correct  if isospin symmetry (more precisely, charge symmetry) is upheld in the nucleon wave function. We  examined these assumption in
\cite{Crawford:2022yhi}, with results, similar to that of an earlier study \cite{Guichon:2011gc}, finding that  there is a correction with the value about equal to the current uncertainty. It is noteworthy that the parton content of nucleons is modified by charge symmetry breaking effects \cite{Londergan:1996vf,Londergan:2005ht,Londergan:2009kj}.  A study of parton charge symmetry breaking is  under current investigation in  JLab  experiment E12-09-002 measuring semi-inclusive $\pi^\pm$ production from the deuteron~\cite{Shia:2022}.

The outline  of this  article  is to first review the work on the nucleon, considering Fermi $\beta$ decay and then discussing how the same charge  symmetry breaking effects in the Hamiltonian are manifest in  
the parton distributions, and how these cause observable effects in  semi-inclusive  pion production on a deuteron target. Thus there is an interesting connection between low and high energy physics.  Finally, the work on nuclear beta decay will be reviewed.

\section{Charge-symmetry breaking  and Fermi $\b$ decay of the neutron}

It has almost always been  assumed that the  quark-level isospin operator is the same as the nucleon isospin operator. This is valid only  if the nucleon wave function is invariant under the specific  isospin rotation known as the charge-symmetry rotation. This operator  rotates an up quark into a down quark, and more formally is a rotation of $\pi$ about the $y$-axis in isospin space~\cite{Miller:1990iz,Miller:2006tv}.
The accuracy of  the implicit assumption was examined long ago by Behrends \& Sirlin~\cite{Behrends:1960nf} who found very small corrections to the beta decay matrix element, on the order of  10$^{-6}$.  This was based on using the neutron-proton mass difference as the perturbing  operator, but this difference does not change the nucleon wave function.
 The work of Ref.~\cite{Crawford:2022yhi} used the non-relativistic quark model to provide a more detailed estimate. An outline of that calculation is presented next.
 
 The non-relativistic quark model \cite{Isgur:1979be} was used, with harmonic oscillator confinement. The Hamiltonian also includes the non-relativistic kinetic energy, the electromagnetic interaction, and the one-gluon exchange interaction. The charge-symmetry breaking  Hamiltonian, $H_1$, originates from the up-down quark mass difference and the fine structure constant, $\alpha$. The effects of charge-symmetry breaking enter only at second and higher orders in $H_1$. Three different quark models were employed. The  differences between models arose from  different estimates of the  strength of the one-gluon exchange hyperfine interaction and the influence of  the pion cloud on the size of the nucleon. 
 
 The matrix element of the isospin operator $\tau_+$ is given to second order by
 \bea
 \la p|\tau_+|n\ra=1 -2 \sum_{k=1} {(p|H_1|p_k)(p_k|H_1|p)\over ({\bar M}-M_k)^2},
 \eea
 where the index $k\ge1$ denotes the $k'$th radial excitation, $\bar M$ is the average of neutron ($n$) and proton ($p$) masses, the rounded brackets represent states computed with the charge symmetry conserving part of the Hamiltonian, and $|p)$ is the proton ground state. Note that the charge-symmetry breaking effects of $H_1$ cause the  matrix element to be less than unity.
 
 In computing the neutron-proton mass difference the  contributions of the up-down quark  mass difference are partially cancelled by the charge-symmetry breaking effects of the kinetic energy, electromagnetic interaction and the one-gluon exchange interaction, leading to the small difference of 1.29 MeV~\cite{Isgur:1979ed}. However, the  up-down quark  mass difference does not change the nucleon wave function, but the other effects act together in concert to modify the wave function. Thus we obtained~\cite{Crawford:2022yhi} a reduction of the $\b$ decay matrix element on the order of 10$^{-4}$, which is  about 100 times larger than the earlier estimate~\cite{Behrends:1960nf}.
 
The schematic calculation of Ref.~\cite{Behrends:1960nf} used second-order perturbation theory in the form of $({\la V\ra\over \D E})^2$, with 
 with $\la V\ra$ taken to be the neutron-proton mass difference and the energy denominator $\D E$ set equal to the nucleon mass.
 The actual value of $ \la V\ra$ must be larger than that for reasons explained above and the energy denominator can be as small as about half the nucleon mass, as determined from  the Roper resonance-nucleon mass difference.
   Thus the early result  was a vast underestimate. Our results of effects of about 10$^{-4}$  (with a spread of about a factor of 5) depend on the details of the quark model, so that the details of the strong interaction model, mentioned above,  enter in important ways. %These details include the strength of the one-gluon exchange hyperfine interaction  and the contribution of the pion cloud to the nucleon size. 
   Nevertheless, it is encouraging  that these results
 are very similar to those of a MIT bag model calculation~\cite{Guichon:2011gc}.  The size of these effects is about the same as the current experimental uncertainties. Future improvements in those uncertainties would increase  the importance of more accurately determining the size of these charge-symmetry violating corrections.

\section{Detecting charge symmetry breaking in the nucleon}
 
Semi-inclusive pion production from the deuteron via lepton deep inelastic scattering, $ed\to e'\pi^\pm +\rm X $, could be a sensitive probe of charge-symmetry breaking effects in nucleon valence parton distributions. \cite{Londergan:1996vf,Londergan:2005ht,Londergan:2009kj}.
For $\pi^+$ production the favored mode of production is from an up quark, while for $\pi^-$ production the favored mode of production is from an down quark. Thus, if charge symmetry is conserved,  the yields, $N_{\rm fav}^{D\pi^\pm}$, obey the relation~\cite{Londergan:1996vf}
\bea
N_{\rm fav}^{D\pi^+}(x,z)=4N_{\rm fav}^{D\pi^-}(x,z),
\eea
where $x$ is the Bjorken variable and $z$ is the ratio of the pion energy to the photon energy.
 The factor of 4 arises from the proportionality of the cross section to the square of quark charges.
The charge-symmetry breaking distributions are defined by 
\bea \d d(x)\equiv d^p(x)-u^n(x);\, \d u(x)\equiv u^p(x)-d^n(x).
\eea
In addition to effects of the charge-symmetry breaking Hamiltonian, $H_1$, there is a kinematic effect of the quark mass  difference that appears in evaluating the probability that a quark carries a light-front momentum fraction $x$.
Calculations of these quantities are reviewed in~\cite{Londergan:1998}, but there is disagreement which of the effects are dominant.
Benesh \& Goldman~\cite{Benesh:1996yg}, in contrast with other authors, find that the kinematic effects of the quark  mass difference are not dominant.
A direct measurement of $\d d$ and $\d u$ would shed light on various aspects of confinement as well as be related to  fundamental $\b$ decay of the neutron.

The ratio~\cite{Londergan:1996vf} 
\bea R^D(x,z)\equiv {4N^{D\pi^-}(x,z)-N^{D\pi^+}(x,z)\over N^{D\pi^+}(x,z)-N^{D\pi^-}(x,z)}
\eea 
is sensitive to the charge-symmetry breaking quantity
\bea
R^D_{CSB}(x)={4\over 3}{\d d(x)-\d u(x)\over u^p_v(x)+d^p_v(x)},
\eea
where the subscript $v$ stands for valence.

Jefferson Laboratory experiment E12-09-02 has taken  data necessary    to obtain measurements of precision ratios of charged pion electroproduction in semi-inclusive deep inelastic scattering from deuterium. Preliminary results indicate that the quantity $\d d(x)-\d u(x)$ can be extracted from the data~\cite{Shia:2022}. A future measurement at an energy-upgraded version of Jefferson Laboratory~\cite{Bodenstein:2022ybi} could be capable of obtaining precision results because of the greater kinematic range available for testing aspects (factorization) of the reaction mechanism. 

\section{Nuclear Superallowed $\b$ decay}

 The dominant contribution to  the unitarity test of the Standard Model  CKM matrix   comes  from the up-down quark matrix element $V_{ud}$.  The value of $V_{ud}$ has been  extracted from nuclear beta decays by Hardy \& Towner in a long series of papers displaying increasing precision culminating in 
~\cite{Hardy:2020qwl}. The highest precision arises from $0^+\to 0^+$ decays from nuclei ranging from $^{10}$C to $^{74}$Rb. The remarkably  consistent nature of the values of $V_{ud}$ obtained from many different decays  led to a very small uncertainty as noted in the introduction. 
 
 Despite the  considerable success of the Hardy \& Towner approach, the  crucial importance of the process in testing the Standard Model  has mandated that the theory behind the analysis be continually re-examined.
 Our focus is on the isospin-breaking correction $\d_C$.  A variation of this quantity, $\D\d_C$  would cause a change in $V_{ud}$ given by
  \be {\D (V_{ud}^2)\over V_{ud}^2}\approx \D\d_C.\label{dv}\ee
 Consider the result $\d_C=0.960(63)$\% for the $0f_{7/2}$ orbital of $^{42}$Ti~\cite{Hardy:2020qwl}. A 20 \% change, for example, in that number is about 0.2\%  and $V_{ud}$ would be changed by half that,  $ 10^{-3}$, a number that is larger than the  current uncertainty.  The Particle Data Group~\cite{ParticleDataGroup:2020ssz} finds an even smaller uncertainty. 
 
  Schwenk \& Miller~\cite{Miller:2008my,Miller:2009cg} raised questions about the Towner-Hardy approach to the isospin correction. We explain their argument.
 In nuclei, the matrix elements of weak vector interactions are not
modified by nuclear forces, except for corrections due to isospin breaking  effects.
Therefore, one has to evaluate the contributions from electromagnetic
and charge-dependent strong interactions to the Fermi matrix element of the isospin raising operator  $\tau_+$,
$M_F =V_{ud} \langle f | \tau_+ | i \rangle$, between the initial and final
states for superallowed $\beta$ decay, $| i \rangle$ and $| f \rangle$. 

Towner and Hardy~\cite{Towner:2007np} use a second quantization formulation
to write the Fermi matrix element as
\be
M_F \approx V_{ud}\la f| w_+|i\ra
,\ee
 with 
 \be w_+=
 \sum_{\alpha, \beta} a_{\alpha}^{\dag} a_{\beta} 
 I^{np}_{\a\b},
\label{MFq}
\ee
where $a_{\alpha}^{\dag}$ creates a neutron in single-particle, shell-model state  with quantum numbers denoted as $\alpha\,(n.l,j,m)$ and 
$a_{\beta}$ annihilates a proton in state $\beta$. The label 
$\alpha$ is used to denote neutron creation and annihilation operators,
while $\beta$ is used for those of the proton. The single-particle matrix element $  I^{np}_{\a\b}$ is given by: 
\be
 I^{np}_{\a\b}
= \delta_{\alpha, \beta} \int_0^{\infty}
R_{\alpha}^n(r) \, R_{\beta}^p(r) \, r^2 dr
\equiv \delta_{\alpha, \beta} \, r_{\alpha} \,,
\label{radi}
\ee
where $R_{\alpha}^n(r)$ and $R_{\beta}^p(r)$ are the neutron and proton
radial wave functions. 
The operator  $w_+$ therefore is seen to act on spatial degrees of freedom, so that it cannot be 
 the same as the isospin operator  $\tau_+$ that  acts {\it only} on isospin degrees of freedom. 
Instead, $ w_+$ of 
\eq{MFq} is the plus component of the W-spin operator of 
MacDonald~\cite{Wspin},  reviewed in Ref.~\cite{DW}.

 Note the appearance of the delta function in \eq{radi}. This means that the radial excitations are missing. In making this approximation Towner \& Hardy  neglected radial excitations of energy  $2\hbar \w$ and higher above the relevant orbitals  and thereby reduced the size of the necessary  shell model space. 
Corrections to the  Towner \& Hardy formalism based on the collective  isovector monopole state were presented in~\cite{Auerbach:2008ut,Auerbach:2021jyt}. Work on the effects of short-ranged correlations appears in~\cite{Lam:2012hp}  that concludes, ``we present a new set of isospin-mixing corrections $\cdots$, different from the values of Towner and Hardy. A more advanced study of these corrections should be performed."

The Towner \& Hardy approach specifically eliminates the influence of short-ranged nucleon-nucleon correlations that involve nucleons in orbitals high above the given shell model space. This strong interaction effect reduces the probability that a decaying nucleon is in a valence single-particle  orbital and suggests that the magnitude of $\d_C$ could be smaller than that of previous calculations.

Condren \& Miller~\cite{Condren:2022dji} argued that the influence of short-ranged correlations between nucleons, unaccounted for by  Towner \& Hardy may  cause  changes in the value of  $\d_C$ that are large on the scale of the desired accuracy.  This means that, depending on future theoretical and experimental work, the value of  $V_{ud}$  could be different than its present value.

Many new experiments  and calculations  have
obtained unambiguous evidence for the existence of  nucleon-nucleon short-ranged correlations  
~\cite{Subedi:2008zz,Fomin:2011ng,Hen:2014nza,Hen:2016kwk,Weiss:2016obx,Stevens:2017orj,Fomin:2017ydn,CiofidegliAtti:2017xtx,Wang:2017odj,Weiss:2018tbu,CLAS:2018xvc,CLAS:2018yvt,Paschalis:2018zkx,Lynn:2019vwp,Ryckebusch:2019oya,Lyu:2019bxr,CLAS:2020mom,CLAS:2020rue,Weiss:2020bkp,Segarra:2020plg,Aumann:2020tcq,Guo:2021zcs} in the time since  Towner \& Hardy started their calculations.
The  significant  effects of short-ranged correlations between nucleons, predicted   long ago, have  been  measured. Such correlations involve the excitations of nucleons to intermediate states of high energy.  Consequently, radial excitations are now known to be important in nuclear physics. 

Spectroscopic factors, essentially
the occupation probability of a single-particle, shell-model orbital, play an important role. As  reviewed in Ref.~\cite{Hen:2016kwk}, electron scattering experiments typically observe only about 60-70\% of the expected number of protons.  This depletion of the spectroscopic factor was observed over a wide range of the periodic table at relatively low-momentum transfer for both valence nucleon knockout using the  $(e,e'p)$  reaction ~\cite{Lapikas:1993uwd} and stripping using the  $(d,^3He)$  reaction \cite{Kramer:2000kc}. The missing strength of 30\%-40\% implies the existence of collective effects (long-range correlations) and short-range correlations in nuclei.
Theoretical analyses~\cite{Atkinson:2019xtc,Geurts:1996zza,Radici:2003zz,Dickhoff:2004xx} made  detailed evaluations finding that including the effects of both long and short-range correlations is necessary to reproduce the results of experiments that measure spectroscopic factors. 
Condren \& Miller argued that, in analogy with the $(e,e'p)$ and $(d,^3He$) reactions, the superallowed beta decay measurements are impacted by the short-ranged correlations that reduce the spectroscopic strength. 

Therefore  Condren \& Miller~\cite{Condren:2022dji} re-examined the calculations of superallowed beta decay rates with an eye toward including the effects of  short-ranged correlations absent in the  Towner \& Hardy formalism.  That work is described here.
 The shell  model, in its simplest form,   places the 
  nucleons  in single particle orbitals, and  the $\beta$ decay matrix element   is simply an overlap between neutron and proton wave functions. If  the Hamiltonian commutes with all components of the isospin operator, the spatial overlap would be unity.
But the non-commuting interactions, such as Coulomb interaction  and the  nucleon mass difference cause the overlap to be less than  than unity. 
This leads to a non-zero value of  the isospin correction known as $\d_C$.   
There must be  a further modification of the value of the matrix element because  the mean field that binds the orbitals is only a first  approximation. The mean-field arises from the average of two- (or more) body interactions,  but residual two- (or more) nucleon effects must remain that cause  long-range correlations,  and those that cause  short-ranged correlations.

The fundamental theory for the Fermi interaction of proton beta decay involves the  isospin operator $\tau_+,$
and the Fermi matrix element is then given by
$
M_F = \langle f | \tau_+ | i \rangle \,, 
$
 $|i\rangle$ and $|f\rangle$ the exact initial and final 
eigenstates of the full Hamiltonian $H=H_0+V_C$, with energy
$E_i$ and $E_f$, respectively and $V_C$ denotes the sum of
{ all} interactions that do not commute with the vector 
isospin operator.

Condren \& Miller  extended the  formalism of Refs.~\cite{Miller:2008my,Miller:2009cg}
by first developing an effective $\b$-decay one-body operator to account for the
dominant isospin-violating effect. The matrix element of this operator was then evaluated in a strongly-correlated system. 
Consider single-particle proton $p$ and neutron $n$ orbitals denoted by $|v,p\ra$ and $|v,n\rangle$, in which the index $v$ denotes the space-spin quantum numbers. These are eigenstate of a Hamiltonian, $h=h_0 +U_C(p)$ with a Coulomb potential, $U_C(p) $ that acts only on  protons.   The eigenkets of $h_0$ are denoted with rounded brackets and those of $h$ with the usual Dirac notation. Then  $|v,n\rangle=|v,n)$.

The use of Wigner-Brillouin perturbation theory in $U_C$ yields the expression:
\bea |v,p\ra=\sqrt{Z_C}|v,p) +{1\over E_v-\L_v h_0\L_v}\L_v U_C|v,p\ra
\label{pw}\eea
with $Z_C= 1-\la v,p| U_C {1\over (E_v-\L_v h_0\L_v)^2} U_C  |v,p\ra$ and the operator $\L_v$ projects out the eigenvalues of $h_0: \,(v,(n,p)|\L_v=0$. 

The single-particle super-allowed beta decay matrix element, $M_{\rm sp}\equiv  (v,n|\t_+|v,p\ra$ is given by:
\bea M_{\rm sp}=V_{ud}\sqrt{Z_C} .  
\eea
 Evaluating $\sqrt{Z_C}$ to second-order in $U_C$ yields
\bea M_{\rm sp}\approx V_{ud}\left(1-{1\over2}(v|U_C {1\over (E_v-\L_v h_0\L_v)^2}\L_v U_C|v)\right). \label{ob}
\eea
  This result repeats the well-known results that the electromagnetic corrections are of second-order.
  The dominant isospin correction of Towner \& Hardy, $\d_C$,    is  twice the second term of the parenthesized expression. 

Next  turn to nuclear super-allowed $\b-$decay.  Define  the one-body Coulomb-correction operator appearing in \eq{ob} as  $\widehat{\cal O}_C(v)\equiv U_C {1\over (E_v-\L_v h_0\L_v)^2}\L_v U_C$.
Consider,  a simplified situation
in which the initial nucleus $i$ consisting of a   proton in a valence orbital $v$ outside an isospin-0 core state of $A$ nucleons  beta decays to a neutron outside the same  state, $f$.  
The core of the state $f$ is taken as that of the state $i$, so that their overlap  does  not influence the $\b$ decay matrix element. Corrections are of  a negligible ${\cal O} (1/A)$.

The dominant  Coulomb correction is obtained from
the matrix element  $\delta_C(v)$ of the operator $\widehat{\cal O}_C(v)$. In coordinate-space and suppressing spin indices, this  is:
\be 
{\d_C}_0(v)=\int d^3rd^3r' \phi^*_v(\bfr) {\cal O}_C(\bfr,\bfr') \phi_v(\bfr').\label{m01}\ee

Condren \& Miller  provided a proof of the validity of \eq{m01} by comparing numerical results with those of exact single particle calculations.
Their result ${\d_C}_0=0.267\,$\%  agrees  with the result  for the $^{46}V$  $f_{7/2}$ state
appearing in  Table I for in Ref.~\cite{Towner:2007np}. 

The simple single-particle state leading to \eq{m01} is only a first, mean-field  approximation to the nuclear wave function. This is because
the valence proton (neutron) undergoes strong interactions with the core nucleons that  involve both
long- and  short-ranged correlations.  Concentrating on short-ranged aspects,
 the two-nucleon wave function is given by
\bea |v,\a\ra=\sqrt{Z_S(v,\a)}|v,\a) +Q  {G\over e}|v,\a),\label{form1}\eea  
 with $\a$ being one of the occupied orbitals of the $T=0$ core state.  The operator  $G$  (the two-nucleon $T$-matrix evaluated at negative energy and modified by Pauli blocking effects)
 is the 
 anti-symmetrized reaction matrix operator  that sums ladder diagrams involving 
two--nucleon interactions. The factor  $Z_S$   insures the normalization,  $e$  represents  an energy denominator and  iterations of the potential that correct the state $|i_0)$ are included  schematically in the  factor $Q {G\over e}$.
The Hermitian projection operator $Q$ obeys $Q|v,\a)=0$,
 and is constructed  to exclude the long-ranged correlations  so that {only} the short-ranged correlations are included in the correction studied here.

  \eq{form1}  is a first step to include the effects of short-ranged correlations.
 Defining $\Omega\equiv Q {G\over e}$, $Z_S(v,\a)=1-(v,\a|\O^\dagger\O|v,\a)$. Then
%\begin{widetext}
 \bea &\d_C(v)=Z_S(v)(v|\OC|v) +  \sum_\a (v,\a|\O^
 \dagger\OC\O|v,\a)\nonumber\\&
. \label{dcv}\eea
%\end{widetext}
with
$Z_S(v)\equiv {}1-\sum_\a (v,\a|\O^\dagger\O|v,\a)	\equiv1- \k(v)
.$ This is the occupation probability, known  to be $<1$. Terms of first-order in $\O$ vanish. 

As a starting point consider that  the existing literature  indicates that 
$\, Z_S(v)\approx 0.8$ for many states $v$, although with dependence on the specific state, nucleus and interactions.  This number comes from many experimental measurements and theoretical calculations cited  above. If one neglects the second term of  the result is that the leading Coulomb correction of TH is multiplied by the factor $Z_S$, potentially
 a very substantial reduction in terms of present accuracy requirements. However, one must include the effects of the second term of
 \eq{dcv}.

Condren \& Miller  estimated  the effects of the  second-order terms in $G$ of \eq{dcv} for the  for the $^{46}V$  $f_{7/2}$ state
appearing in  Table I for in Ref.~\cite{Towner:2007np}. Their estimates were schematic and inconclusive.  The size of the  second-order terms depended on the detailed dependence on the two-nucleon separation distance.  The second-order terms could be negligible or could completely compensate for the reduction caused by having $Z_S<1$.

The  finding was that computations of the isospin correction are strongly sensitive to the effects of short-ranged correlations.  Detailed features of the short-ranged correlations determine whether the influence is an increase, decrease or no change.  The correct  evaluation of this effect can only be assessed precisely by doing detailed calculations using  different models  consistent with  experimentally measured spectroscopic factors. This is important because tests of the unitarity of the CKM matrix demands very high accuracy.  Doing more detailed state-of-art nuclear calculations of superallowed $\b$ decay is therefore a high priority for nuclear theorists.

\section{Discussion}
This paper is concerned with how strong interactions influence the necessary corrections to calculations of $\b$ decay matrix elements caused by the effects of isospin violating interactions. Although of order electromagnetic interactions, strong interaction physics is needed to make  accurate calculations. 

Differences in the details of the 
strong interaction influence the evaluation of the neutron decay matrix element. The size of the  charge-symmetry breaking effects is about the same as the current uncertainties.     Future improvements in those uncertainties would make the importance of these corrections more relevant.
Oddly enough, uncertainties in  the low-energy beta decay matrix element can be constrained by higher-energy  measurements of the  $ed\to e'\pi^\pm +\rm X $ reactions at Jefferson Laboratory.

This paper also discusses the  role of nuclear short-ranged correlations, which affect the value of the correction needed to account for isospin violation in the extraction of the value of $V_{ud}$ from nuclear $\b$ decay.  The effects of short-range correlations have been unambiguously measured and could have a significant effect on the extraction of $V_{ud}$.
 The net result is that  effects previously considered as irrelevant have become relevant. It is noteworthy that acquiring a sufficiently detailed understanding beta decay  will ultimately depend on knowledge of the interplay between the strong, electromagnetic and weak forces.

\acknowledgments{ This research was partially supported by the U.S. Department of Energy Office of Science, Office of Nuclear Physics under Award No. DE-FG02-97ER-41014.
I thank the students Levi Conden and Jacob Crawford for their contributions to the papers that are reviewed here.}

\end{document}